\documentclass[submission,copyright,creativecommons]{eptcs}
\providecommand{\publicationstatus}{To appear in EPTCS.}

\usepackage{amsmath}
\usepackage{amsfonts}
\usepackage{amssymb}
\usepackage{amstext}
\usepackage{epsfig}
\usepackage{graphicx}
\usepackage[all,2cell,knot,poly]{xy}

\newtheorem{example}{Example}

\newcommand{\zb}{\hspace{-0.5pt}}
\newcommand{\zbb}{\hspace{-1.0pt}}

\newcommand{\conct}{\texttt{:}}

\newcommand{\empt}{{\epsilon}}
\newcommand{\Nat}{\mathbb N}

\newcommand{\ddz}{{{\texttt {\footnotesize{\#}}}\zbb}}

\newcommand{\rdot}{{\zb}.{\zb}}

\newcommand{\lsem}{\mbox{$\lbrack\hspace{-0.3ex}\lbrack$}}
\newcommand{\rsem}{\mbox{$\rbrack\hspace{-0.3ex}\rbrack$}}

\newtheorem{proposition}{Proposition}

\newtheorem{definition}{Definition}

\title{Finite Countermodel Based Verification\\
for Program Transformation\\
{\large (A Case Study)}
}

\author{Alexei P. Lisitsa
\institute{Department of Computer Science,\\ The University of Liverpool
}
\email{{a.lisitsa@csc.liv.ac.uk}}
\and
Andrei P. Nemytykh
\institute{Program Systems Institute,\\ Russian Academy of Sciences\thanks{The second author was  supported by RFBR, research project No. 14-07-00133\_a, and 
Russian Academy of Sciences, research project No. 01201354590.
}}
\email{\quad {nemytykh@math.botik.ru}}
}
\def\titlerunning{Finite Countermodel Based Verification for Program Transformation}
\def\authorrunning{A.\,P. Lisitsa, A.\,P. Nemytykh}

\begin{document}
\maketitle

\begin{abstract}
Both automatic
 program verification and program transformation are based on program analysis. In the past decade a number of approaches using various automatic general-purpose program transformation techniques (partial deduction, specialization, supercompilation) for verification of unreachability properties of computing systems were introduced and demonstrated 
 \cite{FioPettPro:01, Hamilton:VPT2015, Leuschel:99, Lis_Nem:IJFCS08}. 
 On the other hand, the semantics based unfold-fold program transformation methods pose themselves diverse kinds of reachability tasks and try to solve them,  aiming at improving the semantics tree of the program being transformed. That means some general-purpose verification methods may be used for strengthening program transformation techniques. This paper considers the question how finite countermodels for safety verification method \cite{Lisitsa:JAR13} might be used in Turchin's supercompilation method. We extract a number of supercompilation sub-algorithms trying to solve  reachability problems and demonstrate use of an external model finder for solving some of the problems.


\end{abstract}

\section{Introduction}\label{sec:Introduction}
A variety of semantic based program transformation techniques commonly called specialization aim 
at improving 
 the properties of the programs based on available information on the context of use. 
The specialization  techniques typically preserve the partial functions implemented by the programs, and given a cost model can be used for optimization.  Specialization can also be used for verification of programs  \cite{FioPettPro:01, Hamilton:VPT2015, Leuschel:99, Lis_Nem:IJFCS08} and for non-trivial transformations, such as compilation  \cite{Ershov:1977,Futamura:1971,Turchin:Report20}. 
Essential for such applications is the interplay between semantic and 
syntactic levels.   When the specialization is applied to a program $P$ it may happen that semantic properties of $P$ are transformed to  simple \emph{syntactic}  properties of   the residual program, which depending on the goals of transformation may be accounted in different ways.

It has been known for a while 
\cite{FioPettPro:01,FioPettProSen:11,Leuschel:04,Turchin:TheoremProving,Tur:86} 
that program transformation techniques, in particular program specialization, can be used to prove some properties of programs automatically. For example, if a program actually implements 
a constant function,  sufficiently powerful and semantics preserving program transformation may reduce the program to a syntactically trivial ``constant'' program, pruning away unreachable branches and proving thereby the property. Viability of such an approach to verification has been demonstrated  in  previous work using supercompilation as a program transformation technique
\cite{Lis_Nem:Programming,Lis_Nem:IJFCS08} 
where it was applied to safety verification of program models of parameterized cache coherence protocols and Petri Nets models \cite{Klimov:PetriNets12,LN08}.
Furthermore,  the functional program modeling and supercompilation have been used to specify and verify 
cryptographic protocols,  and in the case of insecure protocols a supercompiler was utilized in an interactive search for the attacks on the protocols 
 \cite{AhmedLisitsaNemytykh:NSPK,AbdulMSc08,ANepeivoda:PingPong}.
 
 Specialization can be used for analysis of other program properties, based on syntactic properties of the corresponding residual program. For example, any program can be seen as encoding a transition system
 of the parameterized program states and transitions and one may try to apply specialization to  
improve some properties of  the transition graph
(see \cite{Lis_Nem:MissCannVPT2014}).

It is clear that any automated program analysis  can
be applied to program verification. On the other hand, many program analysis tasks, in particular those appearing in the context of program transformation techniques can be considered as verification problems.

In this paper we address the question of applications of a particular verification technique, \emph{finite countermodel based verification (FCM)} 
\cite{GL:08, Lisitsa:RTA12,Lisitsa:JAR13,S01}, 
within the context of \emph{supercompilation}, a  particular program transformation method.  
Supercompilation is a semantic based specialization method introduced by V.~F. Turchin 
\cite{Tur:86} which utilizes in particular unfold-fold based program transformation.  

One of the main challenges in unfold-fold based program transformation techniques is to construct 
a \emph{good}  approximation of the \emph{semantic} tree of the program $P$ being transformed. 
 We want to detect as many unreachable paths in the syntactic tree of $P$ as possible and to prune  away the dead paths.  This syntactic tree is being stepwise unfolded from the program $P$ definition.
Thus, in essence, this challenge is a reachability problem. Supercompilation also poses itself other kind of  reachability tasks and tries to solve them (see Section \ref{sec:GlobalUnreachability} for examples). 

Finite countermodel method (FCM) for safety verification  utilizes the principle of 
\emph{reachability as derivability in the first-order logic} and  reduces the task of safety verification,  even for
infinite state and parameterized systems, to the tasks of \emph{disproving} first-order formulae,  which are tackled then by available finite model finders. 
This principle, proposed initially for the verification of cryptographic protocols in \cite{GL:08,S01}, 
has been later expanded to the larger classes of infinite state and  parameterized verification problems, including safety for general term-rewriting systems \cite{Lisitsa:RTA12}. The method has been shown to be \emph{relatively complete} 
with  respect to more widely known methods of \emph{regular model checking} \cite{Lisitsa:JAR13} and 
\emph{regular tree model checking} \cite{DBLP:journals/corr/abs-1107-5142} and generally it works when the safety can be demonstrated using \emph{regular} invariants (see Section \ref{subsec:Invariants}).

What makes the combination of supercompilation and the FCM method interesting and promising it is their somewhat complimentary features. 
On the one hand, the analysis mechanisms exploited within supercompilation 
do not cover \emph{all} regular properties of parameterized program configurations (sets of states), so any help from the theoretically (relatively) complete FCM method could potentially be useful. On the other hand, supercompilation (as well as other specialization methods) sometimes is able to recognize and use non-regular properties of the program $P$ (see Section \ref{sec:IrregularExamples} for examples) going beyond of what is possible by the FCM method.  

So in short, the main idea of this paper is to explore the  combination of  FCM,  theoretically complete for verification by regular invariants,   with the mechanisms  of supercompilation able to recognize and  deal  with non-regular properties.  

This paper assumes that the reader has basic knowledge of concepts of functional programming, pattern matching, term rewriting systems, and program specialization. 
We also assume that the reader is familiar with the basics of the first-order logic. 

\section{Preliminaries}\label{sec:Preliminaries}

\subsection{The Presentation Language}\label{subsec:PresentationLanguage}

We present our program examples in a language ${\cal L}$ which is a variant of a pseudocode for functional programs, while the supercompilation experiments with the programs were done in the strict functional programming language Refal \cite{Turchin:Refal5}.

The programs given below are written as \emph{strict} (call by value) term rewriting systems based on pattern matching. The rewriting rules (also referred to as sentences) in the programs are ordered from top to bottom and they should be matched in this order.

The data set is a free monoid of concatenation (i.e., the concatenation is associative) with an additional unary constructor, which is denoted only with its parentheses (i.e., without a name).
The colon sign denotes the concatenation. 
The constant $\empt$ is the unit of the concatenation and \emph{may be omitted}, other constants $c$ are
 characters. Let ${\cal C}$ denote a set of the characters.
The monoid ${\cal D}$ of the data may be defined with the following grammar :\\
\texttt{
${\cal D} \ni d$ ::= $\empt$ | $c$ | $d_{1}$\,{\rm :}\,$d_{2}$ | ($d$)
}\\
\noindent Thus a datum 
 in ${\cal D}$
is a finite sequence (including the empty sequence), which can be seen as a forest of \emph{arbitrary} finite trees.

Let  ${\cal F} = \cup_{i}{\cal F}_{i}$ be a finite set of functional symbols, where ${\cal F}_{i}$ is a set of functional symbols of arity $i$. 
%
Let $v, f$ denote a variable and a function name, respectively. The monoid of the corresponding terms may be defined as follows:\\
\texttt{
$t$ ::= $\empt$ | $c$ | $v$ | $f$( $t_{1},\;\ldots{\;},t_{n}$ ) | $t_{1}$\,{\rm :}\,$t_{2}$ | ($t$)
}
, where $n$ is the arity of $f$.

Let the denumerable variable set ${\cal V}$ be disjoined in three
sets ${{\cal V} = {\cal E} \cup {\cal S} \cup {\cal T}}$, where the names from\hfill ${\cal E}$\hfill are\hfill prefixed\hfill with\hfill \texttt{'e{\rdot}'},\hfill while\hfill the names\hfill from ${\cal S}$ \--- with \texttt{'s{\rdot}'}\hfill and\hfill the names\hfill from\hfill ${\cal T}$ \--- with\\ \texttt{'t{\rdot}{\zb}'}. 
{\texttt{s{\rdot}}}{\zbb\zbb\zbb}variables range over 
characters, 
{\texttt{e{\rdot}}}{\zbb\zbb\zbb}variables range over the whole data set ${\cal D}$, 
while a {\texttt{t{\rdot}}}{\zbb\zbb\zbb}variable can take as value any character or any 
data surrounded by parentheses.
For a term $t$ we denote the set of all {\tt e}-variables ({\tt t},{\tt{s}}-variables) in $t$ by ${\cal E}(t)$
 (respectively  
 ${\cal  T}(t)$, ${\cal  S}(t)$). ${\cal V}(t) = {\cal E}(t) \cup {\cal S}(t) \cup {\cal T}(t)$.

Examples of the variables are 
\texttt{s{\rdot}r, t{\rdot}F1, e{\rdot}cls, e{\rdot}memory}.
I.e., a variable name may be any identifier.
We also use a syntactical  sugar for representation of words (finite sequences of characters), so,  for example, the list \texttt{'b'{\rm :}'a'{\rm :}\,$\empt$} is shortened as \texttt{'ba'} and  \texttt{'aba'{\rm :}\,e.x} denotes \texttt{'a'{\rm :}'b'{\rm :}'a'{\rm :}\,e.x}.  

We denote the monoid of terms by ${\cal T(C,V,F)}$. 
A term without function names is said to be \emph{passive}. We denote the set of all passive terms by ${\cal P}({\cal C},{\cal V})$.
Let ${\cal G}({\cal C},{\cal F}) \subset {\cal T}({\cal C},{\cal V},{\cal F})$  
be the set of ground terms, i.e., terms without variables. 
Let ${\cal O}({\cal C}) \subset {\cal G}({\cal C},{\cal F})$ be the set of object terms, 
i.e., ground passive terms.  
Given a subset of the variables 
${\cal V}_1 = {\cal S}_1 \cup {\cal T}_1 \cup {\cal E}_1$ 
where  
${\cal S}_1 \subseteq {\cal S}$, 
${\cal T}_1 \subseteq {\cal T}$, ${\cal E}_1 \subseteq {\cal E}$, a substitution is a mapping $\theta: {\cal V}_1 \rightarrow {\cal T}({\cal C},{\cal V},{\cal F})$ such that 
$\theta({\cal S}_1) \subseteq {\cal C} \cup {\cal S}$ and 
$\theta({\cal T}_1) \subseteq {\cal C} 
\cup {\cal S} 
\cup {\cal T} \cup \{ (t) \; |\; t \in {\cal T(C,V,F)} \}$.  
A substitution can be extended to act on all terms homomorphically.
A substitution is called \emph{object} iff its range is a subset of ${\cal O}({\cal T})$.
We use notation $s=t\theta$ for $s = \theta(t)$, call $s$ an \emph{instance} of $t$ and denote this fact by $s \ll t$. 

A program $P$ 
is a pair $\langle \tau, R \rangle$, where $\tau$ is a 
term called \emph{initial}, and $R$ is a finite list of rewriting rules 
of the form $f(p_1, \ldots, p_k) = r$, where $f \in {\cal F}_k$, for each  $(1 \leq i \leq k)$, $p_i$ is a passive term, $r$ is a term containing the function names only from $R$, ${\cal V}(r) \subseteq {\cal V}(f(p_1, \ldots, p_k))$. 

A program $P = \langle \tau,R \rangle$  with $R = \{l_{i} = r_{i} \mid i = 1\ldots n \}$ gives rise to a reachability relation $\rightarrow_{P}$ on terms as follows.
 For $t_{1}$ and $t_{2}$ the term $t_{2}$ is one-step $P$-reachable from $t_{1}$ if and only if  there exists a substitution $\theta : {\cal V}(t_1) \rightarrow {\cal D}$ such that  1) there exists $i: 1 \le i \le n$ such that for all $j \in \Nat$, $1 \leq j < i$, $t_{1} \theta$ does not match against $l_j$ and it matches against $l_i$, and  2) $t_{2} = r_{i}\theta$. In words, $t_{2}$ is obtained from $t_{1}$ by application of \emph{the first, in the given order} applicable rewriting rule from $R$.   

We denote by $\Rightarrow_{P}$ a one-step  reachability defined similarly to  $\rightarrow_{P}$ above, but omitting the clause ``for all $j \in \Nat$, $1 \leq j < i$, $t_{1} \theta$ does not match against $l_j$''. Thus $t_{1} \Rightarrow_{P} t_{2}$ iff $t_{2}$ is obtained from $t_{1}$  by application of any rule from $R$. 
It is obvious that $\Rightarrow_{P}$ is an \emph{overapproximation} of $\rightarrow_{P}$, that is $\rightarrow_{P}\subseteq \Rightarrow_{P}$.  
We denote \emph{the reflexive, transitive closure} of $\rightarrow_{P}$ and $\Rightarrow_{P}$ by $\rightarrow_{P}^{\ast}$ and $\Rightarrow_{P}^{\ast}$, respectively.  

\subsubsection{Term  Automata, Regular Languages and Invariants}\label{subsec:Invariants}

The following definition is an adaptation of the definition of \emph{forest automata} from \cite{Forest07} to the specific kind  of 
terms  we introduced in Section \ref{subsec:PresentationLanguage}.

\begin{definition}
A term  automaton over a finite set of characters ${\cal C}$ and a finite set of functional symbols   ${\cal F} = \cup_{i}{\cal F}_{i}$  is a tuple ${\cal A} = ((Q,e,\ast), {\cal C}, {\cal F},  \delta_{\cal C}, \Delta_{\cal F},  \delta_{()}, F \subseteq Q)$ 
where  

\begin{itemize}
\item $Q$ is a finite set of states;
\item $(Q,e,\ast)$ is a finite monoid; 
\item $\delta_{\cal C}: {\cal C} \rightarrow Q$; 
\item $\Delta_{\cal F} = \cup_{i} \{\delta_{f} : Q^{i} \rightarrow Q \mid f \in {\cal F}_{i}\}$ is a set of transition functions, one for each $f \in {\cal F}$;
\item $\delta_{()}: Q \rightarrow Q$ is a transition function for  the unary constructor \emph{\texttt{(\;)}};
\item $F$ is a set of accepting states;
\end{itemize}
\end{definition}

 For every \emph{ground}  term  $t \in {\cal T(C,F)}$  the automaton assigns a value $t^{\cal A} \in Q$ which is defined by induction:

\begin{itemize}
\item $\epsilon^{\cal A} = e$;
\item $c^{\cal A} = \delta_{\cal C}(c)$  for $c \in {\cal C}$; 
\item $f(t_{i}, \ldots t_{k})^{\cal A} = \delta_{f}(t^{\cal A}_{1}, \ldots, t^{\cal A}_{k} )$;
\item $t_{1} :  t_{2}^{\cal A} = t_{1}^{\cal A} \ast t_{2}^{\cal A}$;
\item $(t)^{\cal A} = \delta_{()}(t^{\cal A})$ 
\end{itemize} 

A term  $t$ is accepted by the term automaton ${\cal A}$ iff $t^{\cal A} \in F$.
The term language $L_{\cal A}$ of the term automaton ${\cal A}$  is defined as   $L_{\cal A} = \{ t \mid t^{\cal A} \in F\}$.
 A term  language $L$ is called \emph{regular} iff it is a term language $L_{\cal A}$ of some term automaton ${\cal A}$. 

A very general form of unreachability  (safety)   problem we consider in this paper can be specified as follows. 

{\bf Given:}  $R \subseteq {\cal T(C,F)} $, the set of \emph{reachable} terms and $T \subseteq {\cal T(C,F)}$  the set of \emph{target} terms; 

{\bf Question: } Is it true that $R \cap T = \emptyset$?

We say that unreachability  can be established by a \emph{regular invariant}  iff there is regular term language $I$ such that $R \subseteq I$ and $I \cap T = \emptyset$.

\subsection{Examples}

The infinite sequence \texttt{Fib} of Fibonacci words is defined recursively as
\[w_0 = b;\, w_1 = a;\, w_{i+2} = w_i w_{i+1};\]
%
and consists of the words: $b, a, ba, aba, baaba, ababaaba, baabaababaaba, \ldots$

\begin{example}\label{ex:Fib}
The program $\langle \tau, R \rangle$, where $\tau$ is {\tt Fib(e{\rdot}n)} and $R$ given below, computes the $n$-th  pair of consecutive Fibonacci words, where $n$ is given in 
the input argument (the {\tt e{\rdot}n} argument) in the unary notation. The result words are 
separated by the parenthesis constructors rather than the comma sign. For example, 
{\tt Fib('III') = ('aba')\,{\rm :}('baaba')}. 
Note that the right-hand side of the last rule uses associativity of the concatenation {\tt :} the length value of {\tt e.xs} is unknown and it may be greater than $1$.

\noindent
{\tt Fib(e{\rdot}n) = F(e{\rdot}n, 'b', 'a');}\\
{\tt F($\empt$, e{\rdot}xs, e{\rdot}ys) = (e{\rdot}xs)\,{\rm :}(e{\rdot}ys);}\\
{\tt F('I'\,{\rm :}\,e{\rdot}ns, e{\rdot}xs, e{\rdot}ys) = F(e{\rdot}ns, e{\rdot}ys, e{\rdot}xs\,{\rm :}\,e{\rdot}ys);}
\end{example}

The following example demonstrates the use of the associative constructor in the patterns 
(left-hand sides) of the first and second rules. 
The example program defines a predicate {\tt B} testing: 
(1) whether or not the postfix of given Fibonacci word is \texttt{'bb'}; 
(2) given two consecutive Fibonacci words, can the first of them end with \texttt{'b'}, while the second starts with \texttt{'b'}? In the positive case the predicate value is \texttt{'F'}, otherwise it is \texttt{'T'}.

\begin{example}\label{ex:FibTestB}
$\tau$ is {\tt B(Fib(e{\rdot}n))} and $R$ is from the previous example together with:\\
{\tt B( (e{\rdot}xs\,{\rm :}'bb'){\rm :}(e{\rdot}ys) ) = 'F';}\\
{\tt B( (e{\rdot}xs\,{\rm :}'b'){\rm :}('b'\,{\rm :}\,e{\rdot}ys) ) = 'F';}\\
{\tt B( (e{\rdot}xs){\rm :}(e{\rdot}ys) ) = 'T';}
\end{example}

\subsubsection{Pattern Matching}\label{subsubsec:SemanticsPattMatch}
Associativity of the concatenation creates an issue with the pattern matching, namely, given a term $\tau$ and a rule $(l = r) \in R$, then there can be  several substitutions matching $\tau$ against  $l$. 
An example is as follows:  

\begin{example}\label{Example:OpenVars} %
{\tt $\tau =$ f('abcabc', 'bc')} and {\tt $l =$ f(e{\rdot}x\,\conct\,e{\rdot}w\,\conct\,e{\rdot}y, e{\rdot}w)}. There exist two substitutions matching these terms: the first one is $\theta_1({\tt e.x}) = ${\tt 'a'}, $\theta_1({\tt e.w}) = ${\tt 'bc'}, $\theta_1({\tt e.y}) = ${\tt 'abc'}, the second is $\theta_2({\tt e.x}) = ${\tt 'abca'}, $\theta_2({\tt e.w}) = ${\tt 'bc'}, $\theta_2({\tt e.y}) = \empt$. 
\end{example}

To make the pattern matching 
 deterministic
in the presentation language ${\cal L}$, we take the following decision arising from Markov's normal algorithms  \cite{Markov:60} and used in Refal \cite{Turchin:Refal5}: (1) if there is more than one way of assigning values to the variables in the left-hand side of a rule in order to achieve matching, 
then we choose the one in which the leftmost {\tt e}-variable takes the shortest value; (2) if such a choice still gives more than one substitution, then the chosen {\tt e}-variable shortest value is fixed and the case (1) is applied to the leftmost {\tt e}-variable from the {\tt e}-variables excluding considered ones, and so on while the whole list of the {\tt e}-variables in the left-hand side of the rule is not exhausted.

In the sequel we refer to this rule as Markov's rule and such a substitution as Markov's substitution on $l$, matching $\tau$.

\begin{example}\label{Example:OpenVars2} %
{\tt $\tau =$ f('abacad')} and \texttt{$l =$ f(e{\rdot}x\,\conct\,'a'\,\conct\,e{\rdot}y\,\conct\,'a'\,\conct\,e{\rdot}z)}. There exist three substitutions\hfill matching\hfill the terms:\hfill the first\hfill is\hfill $\theta_1({\tt e.x}) =\; $\texttt{$\empt$}, $\theta_1({\tt e.y}) = ${\tt 'b'}, $\theta_1({\tt e.z}) = ${\tt 'cad'};\hfill the second\hfill is\\
 $\theta_2({\tt e.x}) =\; $\texttt{$\empt$}, $\theta_2({\tt e.y}) = ${\tt 'bac'}, $\theta_2({\tt e.z}) = ${\tt 'd'}; the third is $\theta_3({\tt e.x}) = ${\tt 'ab'}, $\theta_3({\tt e.y}) = ${\tt 'c'}, $\theta_3({\tt e.z}) = ${\tt 'd'}. 

The leftmost\hfill {\tt e}-variable\hfill is\hfill {\tt e{\rdot}x}. 
Both\hfill in\hfill the first\hfill and\hfill the second\hfill substitutions\hfill the length\hfill of\hfill the\\
 {\tt e{\rdot}x}'s values is zero. 
The next leftmost {\tt e}-variable is {\tt e{\rdot}y} and 
${\rm length}(${\tt 'b'}$) < {\rm length}(${\tt 'bac'}$)$. The first substitution meets Markov's rule. 
\end{example}

Given a term of the form \texttt{f($t_1,\dots,t_n$)} where for all $(1 \leq i \leq n)$, $t_i \in {\cal O}({\cal C})$ and a term \texttt{f($p_1,\dots,p_n$)} where all $p_i$ are passive terms, the matching \texttt{f($t_1,\dots,t_n$)} against \texttt{f($p_1,\dots,p_n$)} can be viewed as solving the following system of equations in the free monoid of the object terms ${\cal O}({\cal C})$.

\[ 
 \left\{
        \begin{array}{lcl}
              p_1 & = & t_1\\  
                & \ldots & \\  
              p_n & = & t_n  
        \end{array}
 \right. 
\]

We look for all values of the variables (i.e., substitutions $\theta_i$) from ${\cal V}(\texttt{f(}p_1,\dots,p_n\texttt{)})$ such that for each $i$ and each $(1 \leq j \leq n)$, $\theta_i(p_j) = t_j$ and if the values' set is not empty we choose 
the only Markov's substitution, where Markov's rule acts on the pattern $\texttt{(}p_1\texttt{)} \dots \texttt{(}p_n\texttt{)}$. Note the patterns $p_j$ may share variables and this equation system is equivalent to the following single equation $\texttt{(}p_1\texttt{)} \dots \texttt{(}p_n\texttt{)} = \texttt{(}t_1\texttt{)} \dots \texttt{(}t_n\texttt{)}$.
This system has an important property: for all $(1 \le i \le n)\; 
 t_i
\in {\cal O}({\cal C})$. 
There is a simple algorithm solving the equation systems meeting this property.

\subsection{On Supercompilation}\label{subsec:Supercompilation}

In this paper we are interested in one particular approach in program transformation and specialization, known as supercompilation\footnote{From \emph{super}vised \emph{compilation}.}.
Supercompilation is a powerful semantic based program transformation
technique~\cite{SGJ:96,Tur:86} 
having a long history well back to the
1960-70s, when it was proposed by V. Turchin. The main idea behind a supercompiler is to observe the
behavior of a functional program $p$ running on \emph{partially} defined input 
with the aim to define a program, which would be equivalent to the original one (on the domain of the latter), but having improved properties.  
The supercompiler unfolds a potentially infinite tree of all possible computations
of a parameterized program.  In the process, it reduces the redundancy that
could be present in the original program. It folds the tree into a finite
graph of states and transitions  between possible (parameterized)
configurations of the computing system. And, finally, 
it analyses global properties of the graph and
specializes this graph with respect to these properties (without additional
unfolding steps). The resulting program definition is constructed solely based on the
meta-interpretation of the source program rather than by a (step-by-step)
transformation of the program.  

The result of supercompilation may be a specialized version of the original program, taking into account the properties of partially known arguments, or just a re-formulated, and sometimes more efficient, equivalent program (on the domain of the original).

Turchin's ideas have been studied by a number of authors for a long time and
have, to some extent, been brought to the algorithmic and implementation stage
\cite{NT:00}. From the very beginning the development of supercompilation has
been conducted mainly in the context of the programming language Refal
\cite{N:03, Nemytykh:SCP4book, Nem_Pin_Tur, Turchin:Refal5}
based on syntax and semantics similar to that of our presentation language ${\cal L}$.  
%
A number of 
model 
supercompilers for subsets of functional languages based on Lisp data were implemented with the aim of formalizing some aspects of the supercompilation algorithms 
\cite{Jonsson_Nordlander, Klyuchnikov:HOSC, Mitchel_Runciman, SGJ:96}. 
The most advanced supercompiler for Refal is SCP4 
\cite{N:03, Nemytykh:SCP4book, NT:00}.

\section{Finite Countermodels and Program-State Reachability}\label{sec:FinCountermodReachSt}

Given a term $t \in {\cal T}({\cal C},{\cal V},{\cal F})$, the set of instances of $t\theta$ such that the substitution $\theta$ is 
an object on ${\cal V}(t)$, ranging over ${\cal O}({\cal C})$,
is called \emph{the state set} of $t$. 
Given a program $P = \langle \tau, R \rangle$ in ${\cal L}$ 
(see Section \ref{subsec:PresentationLanguage}), 
the state set $S_0$ of $\tau$ is called \emph{the initial state set} of $P$. 
%
The rewriting system $R$ is able to evolve according to relation $\rightarrow_{P}$, starting from a fixed state $s_0 \in S_0$ and producing some more states of $P$.
Suppose that $S_0$ is described by a predicate (characteristic function) $\Sigma_0(\cdot)$ written in a logical language ${\cal M}$.
Let $\cal{B}$ be a formal theory defined in $\cal{M}$ and $\phi(\cdot)$ be a formula in $\cal{M}$, describing some state set of $P$.
Assume that $R$ satisfies the following: given two states $s_0$, $s$ of $P$, 
if $s$ is reachable  via $\rightarrow_{P}^{\ast}$ from $s_0$ then ${\cal B},\phi(s_0) \vdash \phi(s)$. 
Suppose that a formula $\psi(\cdot)$ (hypothesis in ${\cal B}$) specifies a set of bad states, 
 whose reachability contradicts a safety property of the program $P$. 
 Then refutation of $\psi(s)$ (in the theory ${\cal B} \wedge \phi(s_0)$) will mean the fact of safety of $P$ \--- unreachability of the states satisfying the formula $\psi(\cdot)$. One may refute the hypothesis $\psi(s)$ by producing
a countermodel of the theory 
${\cal B} \wedge \phi(s) \rightarrow \psi(s)$.

\subsection{The Formal Theory of ${\cal D}$}\label{subsec:FormalTheoryD}

Let us redefine the data monoid ${\cal D}$ very slightly by providing an explicit name for the additional free unary operation. 
Henceforth, we assume that the characters set ${\cal  C}$ of the language ${\cal L}$ is finite and $\beta,\gamma \notin {\cal C}$, ${\texttt :} \notin {\cal C}$.
Let $\beta$ stand for the unary operation. In this encoding the data set ${\cal D}$ is redefined as follows:\\
\texttt{
${\cal D} \ni d$ ::= $\empt$ | $\gamma$ | $d_{1}$\,:\,$d_{2}$ | $\beta(d)$
} 
, where $\gamma$ ranges over ${\cal C} \setminus \{ \empt \}$.

Let ${\cal C}$ be $ \{ \empt,  \texttt{'a'}, \texttt{'b'}\}$, 
consider the following theory $T_{\cal D}$ in the first-order predicate logic:
\begin{flushleft}
$\forall x,y,z{\;}.\; (x : y) : z = x : (y : z) $\\
$\forall x{\;}.\; x : \empt = x $\\
$\forall x{\;}.{\;} \empt : x = x$\\
$(\neg (\empt = \texttt{'a'})) \wedge (\neg (\empt = \texttt{'b'})) \wedge (\neg (\texttt{'a'} = \texttt{'b'}))$\\
$R(\empt) \wedge R(\texttt{'a'}) \wedge R(\texttt{'b'})$\\
$\forall x{\;}.\; R(x) \rightarrow R(\beta(x))$\\
$\forall x,y{\;}.\; R(x) \wedge R(y) \rightarrow R(x : y)$
\end{flushleft}

The first three axioms are the free monoid axioms: the first one expresses associativity of the concatenation, the second and third axioms say the constant $\empt$ is the identity element. 
The last three axioms axiomatize the unary predicate $R(\cdot)$ reflecting the inductive definition of the data set.

The theory $T_{\cal D}$ represents the data set ${\cal D}$ as stated in the following proposition 

\begin{proposition}
$d \in {\cal D} \Leftrightarrow T_{\cal D} \vdash R(d)$
\end{proposition}

\section{Unfolding and ${\cal L}$-Program-State Reachability}\label{sec:UnfoldingReachSt}

Let us briefly recall some basic concepts of program specialization which we need below. More details can be found in \cite{Mitchel_Runciman, Nemytykh:SCP4book, SGJ:96, Turchin:Report20, Tur:86}.

Given a function call $\texttt{st}_0 = \texttt{f}_k\texttt{(d}_0\texttt{)}$, 
where 
$\texttt{f}_k \in {\cal F}_{k},\; \texttt{d}_0 \in {\cal D}^k$, the abstract ${\cal L}$-machine \texttt{Int(p,$\cdot$)}, starting from the state $\texttt{f}_k\texttt{(d}_0\texttt{)}$ by matching $\texttt{f}_k\texttt{(d}_0\texttt{)}$ against the left-hand sides $l_i$ 
of the rules defining $\texttt{f}_k$,  chooses a particular rule $\rho_{i_0}$ of $\texttt{f}_k$ and constructs a next state based on the right-hand side of $\rho_{i_0}$. This matching algorithm can be seen as an algorithm 
 solving 
the following  equations $l_i = \texttt{f}_k\texttt{(d}_0\texttt{)}$. 
%
That is to say, the matching chooses the Markov's substitution
  $\sigma(\cdot): {\cal V}(l_i) \rightarrow {\cal D}$ such that 
  $\sigma(l_i) = \texttt{f}_k\texttt{(d}_0\texttt{)}$,
  if such a mapping exists. 
  Let  \texttt{Step(st$_0$)} denote the result of applying an algorithm including the successful pattern matching and the replacement of \texttt{st$_0$} with $\sigma(r_i)$. 

In general, \texttt{Int(p,$\cdot$)} 
 iterates the execution of \texttt{Step(st)} 
when the state \texttt{st} is a configuration from the function stack top and the input data of the state \texttt{st} are completely known. 

The unfolding algorithm approximates the semantic tree of \texttt{p} by means of iterating a meta-extension \texttt{MStep($\cdot$)} of \texttt{Step($\cdot$)} 
in the case when the state \texttt{st} is partially unknown. 
  The execution of   \texttt{MStep(st)} results in a tree whose branches correspond to the subsets  of the input parameters values. 
The branchings are produced by a meta-extension of the matching, i.e., by an algorithm solving the equations $l_i = \texttt{st}$ of a general form in the term monoid ${\cal T}({\cal{C}},{\cal{V}},{\cal{F}})$. 
In the sequel we refer to this meta-extension as the extended pattern matching.
In particular, the algorithm has to solve word equations: 
in general, this task is nontrivial 
(see \cite{Jaffar:90, Khmelevskii:71, Makanin:77, Plandowski:2006}), 
although there exist several simple algorithms for solving such 
equations when they are of some restricted forms 
\cite{Dabr_Pland:04, Diekert02, Jez:13}.

\subsection{One-Step Unreachability}\label{subsec:One-stepUnreachability}

Taking into account the hardness of Makanin's algorithm \cite{Makanin:77} solving word equations of general forms, one may approximate the semantic tree as follows. Let a program rule $l = r$ and a parameterized state \texttt{st} be given. 
Before executing the algorithm \texttt{F} for solving the equation 
$l = \texttt{st}$, \texttt{MStep} tries to prove that this equation has no solutions, 
 using the algorithm  
\texttt{NoSol}, 
not necessarily complete for word equations.

If 
 \texttt{NoSol}
does not finish its work in the given time limit, we say that  
 \texttt{NoSol}
  fails in proving the unsatisfiability. In the fail case we just call \texttt{F}. 
 If 
  \texttt{NoSol}
 succeeds, then the program rule $l = r$ being considered is unreachable from the parameterized state \texttt{st} and the tree branch corresponding to this rule is pruned away. If \texttt{F} succeeds, then it, like a Prolog interpreter, may return a simple narrowing of the parameters of \texttt{st} and Markov's substitution depending on the narrowed parameters and satisfying the equation $l = \texttt{st}$. This substitution allows us to proceed with the unfolding. 
It may happen though that the narrowing of the parameters is not expressible in the language ${\cal U}$ describing the parameterized configurations of the program being transformed.
In particular, the set of solutions of such  an equation may require a recursion for its definition, while the language ${\cal U}$ may lack recursion and iteration constructions.
We now exemplify the situation.
The finite countemodel finding is used as an  
 \texttt{NoSol}
procedure.

\begin{example}\label{ex:RepeatedE-Vars}
Consider the following program 
$p = \langle \tau, R \rangle$, $\tau =\; ${\tt f('a'\,\conct\,e{\rdot}q, e{\rdot}q\,\conct\,'a')} and $R$ is
\begin{flushleft}
{\tt f(e{\rdot}x, e{\rdot}x) = 'T';}\\
{\tt f(e{\rdot}x, e{\rdot}y) = 'F'\,\conct\,(e{\rdot}x)\,\conct\,(e{\rdot}y);}
\end{flushleft}

The extended pattern matching has to solve the following system of the parameterized equations:
\[ 
 \left\{
        \begin{array}{lclr}
              {\tt e.x} & = & {\texttt {\tt 'a'\,\conct\,e{\rdot}q}} & \\  
              {\tt e.x} & = & {\texttt {\tt e{\rdot}q\,\conct\,'a'}} & \mbox{\hspace{50pt}{\LARGE $^{^{(\star)}}$}}
        \end{array}
 \right. 
\]
It is equivalent to the single relation $\Phi({\tt e.q})$ (equation) \--- {\tt 'a'\,\conct\,e{\rdot}q $=$ e{\rdot}q\,\conct\,'a'} imposed on the parameter {\tt e{\rdot}q}.\footnote{It is easy to see that its solution set is $\{\theta_i({\tt e.q}) =\; ${\tt 'a'}$^i\; |\; i \in \Nat \}$.} 

At the first glace, $\Phi({\tt e.q})$ must be the predicate labeling the first branch 
coming out of
the semantics tree root, while the second branch must be labeled by its negation 
$\neg\Phi({\tt e.q})$. 
$\Phi({\tt e.q})$ narrows the range of {\tt e{\rdot}q}. 
But the problem is that $\Phi({\tt e.q})$ cannot be represented in the pattern language, using at most finitely many 
patterns to define 
the program result of the unfolding. Recursion should be used to check whether or not a given input data belongs to the truth set of $\Phi({\tt e.q})$. 

The {\tt e{\rdot}x}-variable multiplicity $\mu_{{\tt e.x}}({\tt f(e.x, e.x)}) > 1$ causes this problem: the system $(\star)$ implies an equation, where the parameter {\tt e{\rdot}q}\hspace{3pt} plays the role of a variable and\hspace{3pt} \emph{both sides of the equation}\hspace{3pt} contain {\tt e{\rdot}q}.
\end{example}

Let us replace the initial parameterized configuration in Example \ref{ex:RepeatedE-Vars}: 
\[\tau = \texttt{{\tt f('a'\,\conct\,e{\rdot}q\,\conct\,'a'\,\conct\,e{\rdot}q\,\conct\,'b', e{\rdot}q\,\conct\,'a'\,\conct\,e{\rdot}q\,\conct\,'b'\,\conct\,e{\rdot}q)}}\] 

Now the extended pattern matching has to solve the following equation 
\[\texttt{{\tt 'a'\conct\,e{\rdot}q\,\conct'a'\conct\,e{\rdot}q\,\conct'b'} =  \texttt{e{\rdot}q\,\conct'a'\conct\,e{\rdot}q\,\conct'b'\conct\,e{\rdot}q}}\]
\noindent 
It is easy to see that this equation having variable {\tt e{\rdot}q} both in 
 the left and right-hand sides is inconsistent in ${\cal D}$.
Mace4 automated finite model finder by W.\,McCune \cite{Mace4} quickly recognizes this fact in the context of the first-order theory 
$T_{\cal D}$ (see Section \ref{subsec:FormalTheoryD}). I.e., Mace4 finds a finite countermodel of the following formula 
$
\exists \texttt{e{\rdot}q}\;.\, (\texttt{{\tt 'a'\conct\,e{\rdot}q\,\conct'a'\conct\,e{\rdot}q\,\conct'b'} =  \texttt{e{\rdot}q\,\conct'a'\conct\,e{\rdot}q\,\conct'b'\conct\,e{\rdot}q}})
$
in the theory of ${\cal D}$.
That is, unfolding the program being considered in the given context 
can prune away the first rule of $R$ and result in the following program 
(which can be viewed as a tree):\\
$p_1 = \langle \tau_1, R_1 \rangle$, $\tau_1 = $
\texttt{{\tt f$_1$('a'\conct\,e{\rdot}q\,\conct'a'\conct\,e{\rdot}q\,\conct'b', e{\rdot}q\,\conct'a'\conct\,e{\rdot}q\,\conct'b'\conct\,e{\rdot}q)}}
and $R_1$ is the only rule:
%
{\tt f$_1$(e{\rdot}x, e{\rdot}y) = 'F'\conct(e{\rdot}x)\conct(e{\rdot}y);}

Note that we did not construct any narrowing of the parameter \texttt{e{\rdot}q} and the information on the property of \texttt{e{\rdot}q} (i.e., the equation above is inconsistent) is lost. The constructed tree is an approximation of the semantic tree of $\langle \tau, R \rangle$, rather than the exact semantics tree. If the right-hand side of the remaining rule includes a function call then, in general, the lost information might be useful for further specialization. For example, it might be \texttt{'F'\,\conct\,g((e{\rdot}x)\,\conct\,(e{\rdot}y))}.
The configuration $\tau_1$ might be encountered in an internal vertex of the unfolding tree.

The example above demonstrates a potential feature of using a finite countermodel finder for improving the approximation of the semantics tree generated by the unfolding. 
An interface linking the supercompiler SCP4 \cite{N:03, Nemytykh:SCP4book, NT:00} with Mace4 has been implemented. 
It allows formulating in Mace4 such a kind of unreachability problem and returning to SCP4 
the result produced by Mace4 during a time limit indicated by a user.

At first glance, the construction given in Example \ref{ex:RepeatedE-Vars} can be generalized as follows. Let a program $P = \langle t, R \rangle$, $t = f(u_1,\ldots,u_k)$ and $R$ \---  
a function $f$ definition below, 
where $k \in \Nat, f \in {\cal F}_k$ and for all $(1 \leq i \leq k)$ $u_i \in {\cal P}({\cal C},{\cal V})$, be given. 
Let $\ddz{\cal V}(t) = m \in \Nat$ and $\ddz{\cal V}(l_i) = s_i \in \Nat$. 
Let the sets ${\cal V}(t), {\cal V}(l_1),\ldots, {\cal V}(l_n)$ be ordered. 
Let $q_j$ stand for the $j$-th element of ${\cal V}(t)$, while $w_{ij}$ stand for the $j$-th element of ${\cal V}(l_i)$.

\[ 
 \left\{
        \begin{array}{lcl}
              f(p_{11},\dots,p_{1k}) & = & r_1\\  
                & \ldots & \\  
              f(p_{n1},\dots,p_{nk}) & = & r_n  
        \end{array}
 \right. 
\]

\begin{definition}\label{def:Reachability} Given $i \in \Nat$, $1 \leq i \leq n$, the rule $l_i = r_i$ of the function $f$ is said to be one-step reachable from the term $t$ if there exists a substitution $\theta : {\cal V}(t) \rightarrow {\cal D}$ such that for all $j \in \Nat$, $1 \leq j < i$, $t\theta$ does not match against $l_j$ and it matches against $l_i$. A rule is said to be one-step  unreachable if it is not one-step reachable.
\end{definition}

Let $i \in \Nat$, $1 \leq i \leq n$, be given. 
One may suspect  
that refuting the following formula, expressing reachability 
of a rule $l_i = r_i$,  leads to proving that the rule  
of the function $f$ above is one-step unreachable from the term $t$.\\
$
\exists \texttt{e{\rdot}v}\, \exists q_1,\ldots, q_m\,.\, 
                        ((\texttt{e{\rdot}v} = \texttt{(}u_1\texttt{)}\ldots\texttt{(}u_k\texttt{)}) \wedge 
                             (\forall w_{11},\ldots,w_{1s_1} 
                             \neg(\texttt{e{\rdot}v} = \texttt{(}p_{11}\texttt{)}\ldots\texttt{(}p_{1k}\texttt{)}
                                         )) \wedge\\
                     \texttt{\hspace{250pt}}\ldots\\ 
                     \texttt{\hspace{188pt}}
                                    (\forall w_{(i-1)1},\ldots,w_{(i-1)s_{(i-1)}} 
                                    \neg(\texttt{e{\rdot}v} = \texttt{(}p_{(i-1)1}\texttt{)}\ldots\texttt{(}p_{(i-1)k}\texttt{)}
                                         ))\wedge \\
                     \texttt{\hspace{188pt}}
                            (\exists w_{i1},\ldots,w_{is_{i}} 
                            (\texttt{e{\rdot}v} = \texttt{(}p_{i1}\texttt{)}\ldots\texttt{(}p_{ik}\texttt{)}
                                         )) \\
                     \texttt{\hspace{80pt}}
                     )
$\\
%
\noindent
But the variables here are assumed to range over the data set ${\cal D}$ only  and so the na{\"{\i}}ve  application of a generic finite model finder may lead to vacuous countremodels, refuting the formula  on  a domain of elements unrelated to ${\cal D}$.  
One may still try to analyse such conditions automatically, possibly using alternating applications of the first-order model finder and a theorem  prover. We will address this issue elsewhere.

Nevertheless we can overapproximate the reachability condition.
Namely, we do not 
 consider
any rule excluding the current rule being explored. That is to say, we approximate the ${\cal L}$-pattern matching with the pattern matching used in non-deterministic term rewriting. The corresponding formula to be refuted is as follows:
$
\exists q_1,\ldots, q_m\,.\, \exists w_{i1},\ldots,w_{is_{i}}\,.\,
   \texttt{(}u_1\texttt{)}\ldots\texttt{(}u_k\texttt{)} =  
      \texttt{(}p_{i1}\texttt{)}\ldots\texttt{(}p_{ik}\texttt{)}$.

So the refutation of this formula proves  unreachability by non-deterministic rewriting and therefore original  unreachability.

Concluding this section we emphasize that the worst-case time complexity of the program resulting
in Example \ref{ex:RepeatedE-Vars} is ${\cal O}(1)$, while the worst-case time complexity of the original program is ${\cal O}(n)$, where $n$ is the input data size. 
  Improving this complexity was possible by the use of Mace4, which eliminated a rule of the original program with a hidden loop over the input data. 

\section{Global Unreachability}\label{sec:GlobalUnreachability}

Unlike the most known specialization techniques supercompilation may extend the domain of the partial function defined by the program being transformed. That makes supercompilation more flexible as compared with those methods. For example, supercompilation is able to improve the worst-case time complexity of some input programs, while partial evaluation cannot \cite{Jones:Complex}.
Other transformation techniques such as distillation  \cite{Hamilton:09} can also improve the worst-case time complexity of some programs.
 In this section we consider one of the supercompilation tools for such a kind of  transformations 
 assisted by the finite countermodel method.

\subsection{Online Generated Program Output Formats}\label{subsec:OutputFormats}

Given a program $P = \langle t, R \rangle$ and a substitution $\theta : {\cal V}(t) \rightarrow {\cal D}$, by $\lsem t\theta \rsem$ we denote the result of a (finite) computation 
of $t\theta$ according to the program $P$. 

\begin{definition}\label{def:OutputFormats} 
Let a program $P = \langle t, R \rangle$ be given. 
A term $u \in {\cal P}({\cal C},{\cal V})$ is said to be an output format of the program $P$ if for any substitution $\theta : {\cal V}(t) \rightarrow {\cal D}$ there exists a substitution 
$\eta : {\cal V}(u) \rightarrow {\cal D}$ such that $u\eta = \lsem t\theta \rsem$.
Let $u_1$ and $u_2$ be two output formats of $P$. 
If $u_1 \ll u_2$, then we say $u_1$ lesser than $u_2$. If $u_1$ lesser than any other output format of $P$, then $u_1$ is said to be a minimal output format of $P$. 
\end{definition}

The minimal\hfill output\hfill format\hfill of\hfill a given\hfill program\hfill is\hfill not\hfill unique. 
Examples\hfill of\hfill the output\hfill formats\hfill are:\\
 \texttt{e{\rdot}x} is an output format of any program; 
\texttt{s{\rdot}y} is the only (modulo variable renaming) minimal output format of the program defined in Example \ref{ex:FibTestB};
both \texttt{s{\rdot}z\,\conct\,e{\rdot}x} and \texttt{s{\rdot}z\,\conct\,e{\rdot}x\,\conct\,e{\rdot}y} are minimal output formats of the program given in Example \ref{ex:RepeatedE-Vars}.

The tree $T$ being stepwise generated by unfolding is potentially infinite. As a consequence it is an object to be somehow folded back into a finite graph representing the residual program $Q$. The folding algorithm works stepwise online, i.e., given an intermediate state of $T$ the algorithm tries to fold a potentially infinite path in this intermediate state into a loop, using generalization of parameterized
 configurations (states)  of the original program $P$. We omit the details of the generalization algorithm. 
 Edges folding such paths are called \emph{references}. 
 Thus the intermediate state of $T$ actually is a graph $G$ rather than a tree (see Fig. \ref{UnfoldGraph}).

Given a vertex (a parameterized state of $P$) $v$ of $G$, if all references from the vertices on the paths originating in $v$ are incoming in the vertices from the same path set, then the corresponding part of $G$ rooted in $v$ is a self-sufficient (closed) subgraph. 
Such a vertex $v$ is called the root of the subgraph. The root is a potential entry point into one of the residual functions (sub-programs), i.e., the root is an input format of a residual function $H$. Let such a root just be created by a step of the folding algorithm. The supercompiler SCP4 analyses the subgraph $H$ and constructs its output format.
The calls of the folding algorithm are stepwise interleaved with the calls of the unfolding algorithm. Hence $G$ may include both some completely folded subgraphs and still non-unfolded parameterized configurations of $P$. Such configurations may include some calls of the residual function $H$. 
A non-trivial output format\footnote{The trivial output formats are \texttt{e{\rdot}x}, \texttt{e{\rdot}x\conct{e{\rdot}y}}, \texttt{e{\rdot}x\conct{e{\rdot}y}\conct{e{\rdot}z}} and so on.} 
of $H$ restricts $H$'s image set. Therefore the information brought beyond $H$'s recursions (loops) might be used for specializing the function call (e.g., \texttt{g(e{\rdot}z)} in Fig. \ref{UnfoldGraph}) using the call of $H$ as an argument. That immediately implies the worst-case time complexity of the residual program $Q$ may be reduced as compared with the worst-case time complexity of the original program $P$.

Note that the non-trivial output format may be a property of the original program $P$ 
in the context of specialization (the initial configuration). 
In general, for example, $P$ or its configuration being considered \textit{per se} may have only the trivial output format. Thus the \emph{online} generating of the output formats does matter.

{
\begin{figure}
\[
\def\objectstyle{\scriptstyle} 
\def\labelstyle{\scriptstyle}
\xymatrix @+2mm @R-15pt @C-20pt {
  & *++[F-:<3pt>]{\texttt{let e.z = f(e.x) in g(e.z)}}
     \ar[d]^{} 
     \ar[drr]^{} 
     &    &
\\
 &  *++[F-:<3pt>]{\texttt{v: f(e.x)}} 
     \ar[d]^{} 
     \ar[dr]^{} 
     &    & *++[F..:<3pt>]{\texttt{v$_1$: g(e.z)}}
\\
 &  *++[F-:<3pt>]{\mathbf{\bullet}} 
     \ar[d]^{} 
     \ar[dl]^{} 
     &  *++[F=:<3pt>]{\mathbf{\clubsuit}}  &
\\
 *++[F=:<3pt>]{\mathbf{\clubsuit}}
     &  *++[F-:<3pt>]{\texttt{}} 
        \ar[dr]^{} 
        \ar[dl]^{} 
     &    &
\\
 *++[o][F]{\texttt{\hspace{15pt}}}
              \ar@/^8pc/@{-->}[uuur]^<{} 
     &    & *++[F-:<3pt>]{}
              \ar[dl]^{} 
              \ar[dr]^{} 
     &    \vspace{-100pt}
\\ 
     & *++[F=:<3pt>]{\mathbf{\clubsuit}} 
     &    & *++[o][F]{\texttt{\hspace{15pt}}}\ar@/^-2.5pc/@{-->}[uuull]{..} 
}
\]
\caption{An intermediate state of an unfold-fold graph $G$: (1) the subgraph $H$ rooted in \texttt{v} is self-sufficient, while (2) the subgraph rooted in $\bullet$ is not. The configuration \texttt{v$_1$} still is not unfolded.}\label{UnfoldGraph}
\end{figure}
}

The output format of the subgraph $H$ is constructed by generalizing the formal exits from the recursions defining $H$ (the double-boxing $\clubsuit$ leaves in Fig. \ref{UnfoldGraph}). Hence the lesser number of such exits the subgraph $H$ includes the more specific output format of $H$ may be constructed (see Definition \ref{def:OutputFormats} above).
The recursion exits are presented as $H$'s edges belonging to the paths outgoing from the root $v$. 
Thus the following problem arises.  
Given a subgraph $H$ representing a potential residual sub-program one has to prove unreachability of 
as many  as possible syntactic recursion exits. 

It is exactly the task we propose to delegate to a finite countermodel finder.
Concrete executions of the finder may take too long times and even may not terminate. We suggest using the finder as explained in Section \ref{subsec:One-stepUnreachability}.

A formal theory for the image set (or its superset) of the partial function $H$ needed to call the finder might be constructed by a compiler from the term rewriting language ${\cal L}$ into 
 the first-order logic
language. The theory and a goal hypothesis to be refuted by the finder should somehow be based on the syntax of the subgraph corresponding to $H$.

For the reasons explained in Section 
\ref{subsec:One-stepUnreachability} 
we can overapproximate
the ${\cal L}$-unreachability with the non-deterministic term rewriting unreachability.

The simplest hypothesis is the assumption that several given syntactic exit-branches from the recursion defined by $H$ are unreachable from the parameterized state in the root of the subgraph.
Below we consider simple examples of such formal theories and goals.

\subsection{Two Very Simple Output Formats}\label{subsec:SimplestOutputFormats}

A very simple 
hypothesis is that the partial (sub)function being analyzed 
is \emph{not} the empty partial function. 
I.e., at least one of its syntactic exits is reachable from the state in the root of the subgraph. Refuting this hypothesis means the subgraph root itself is unreachable from the root of the tree $T$ and the branch leading in this subgraph root is dead. Any term in ${\cal P}({\cal C},{\cal V})$ is an output format of the empty partial function.

If the analysis of the hypothesis above does not lead to the empty partial function, 
the following hypothesis could be made: the output format is a datum $d \in {\cal D}$. 
I.e., the (super)image set of a (sub)function being analyzed includes the single datum $d$. 
We might prove this fact if we are able to refute reachability of all syntactic exits excluding one, which 
 returns the datum $d$.

Let us consider the program $P$ given in Example \ref{ex:FibTestB}. If the first two rules of the function \texttt{B} are unreachable from the initial configuration $\tau = \texttt{B(Fib(e{\rdot}n))}$, then the minimal output format of $P$ is \texttt{'T'}. The supercompiler SCP4 is able to prove this fact by itself \--- without any call of a countermodel finder. This fact directly implies that none of the Fibonacci words contains \texttt{'bb'} as a subword.
Let us slightly change the predicate given in Example \ref{ex:FibTestB}  as follows.

\begin{example}\label{ex:FibTestA}
Let us consider $\tau$ to be {\tt A(Fib(e{\rdot}n))} and $R$ to be from Example \ref{ex:Fib} together with:\\
{\tt A( (e{\rdot}xs\,{\rm :}'aaa')\,{\rm :}(e{\rdot}ys) ) = 'F';}\\
{\tt A( (e{\rdot}xs\,{\rm :}'aa')\,{\rm :}('a'\,{\rm :}\,e{\rdot}ys) ) = 'F';}\\
{\tt A( (e{\rdot}xs)\,{\rm :}(e{\rdot}ys) ) = 'T';}
\end{example}

SCP4 proves unreachability of 
the first two of the rules for \texttt{A} 
from the configuration $\texttt{A(Fib(e.n))}$ and generates the minimal output format \texttt{'T'}. 
That means none of the Fibonacci words contains \texttt{'aaa'} as a subword. 
A more natural encoding of the same problem is presented 
in Example \ref{ex:FibTestOpen}. 
For that encoding  SCP4 fails to prove both 
 properties of  the Fibonacci words. 
It cannot recognize that the first rules of both \texttt{A} and \texttt{B} are unreachable from the corresponding initial configurations.

\begin{example}\label{ex:FibTestOpen}\texttt{\hspace{5pt}}\\
{\tt A( (e{\rdot}xs)\,{\rm :}(e{\rdot}ys\,{\rm :}'aaa'\,{\rm :}\,e{\rdot}zs) ) = 'F';}\\
{\tt A( (e{\rdot}xs)\,{\rm :}(e{\rdot}ys) ) = 'T';}\\
{\tt B( (e{\rdot}xs)\,{\rm :}(e{\rdot}ys\,{\rm :}'bb'\,{\rm :}\,e{\rdot}zs) ) = 'F';}\\
{\tt B( (e{\rdot}xs)\,{\rm :}(e{\rdot}ys) ) = 'T';}
\end{example}

One may try to prove that those first two rules are unreachable from $\tau$, using the finite coutermodel method (see Section \ref{sec:FinCountermodReachSt}).
For the reasons explained in Section \ref{subsec:One-stepUnreachability} we have to overapproximate the ${\cal L}$-reachability $\rightarrow^{\ast}$ with the non-deterministic term rewriting reachability $\Rightarrow^{\ast}$. A first-order theory has to be 
 generated,
in which derivability is compatible with 
the overapproximated  reachability condition in the program being considered. 

Following \cite{Lisitsa:TalkCachan12} we here consider a simpler example of a first-order theory $Fib_0$ demonstrating how to establish similar properties automatically using first-order theorem disproving by finite countermodels finding. The theory $Fib_0$ is as follows:

\vspace{5pt}
\noindent
$
T_{\cal D}\\
K(\texttt{\tt 'b'}, \texttt{\tt 'a'}).\\
K(\texttt{\tt e{\rdot}xs}, \texttt{\tt e{\rdot}ys}) \rightarrow K(\texttt{\tt e{\rdot}ys}, \texttt{\tt e{\rdot}xs}\,{\rm :}\,\texttt{\tt e{\rdot}ys}).\\
A(\texttt{\tt e{\rdot}ys\,{\rm :}'aaa'\,{\rm :}\,e{\rdot}zs}).\\
B(\texttt{\tt e{\rdot}ys\,{\rm :}'bb'\,{\rm :}\,e{\rdot}zs}).
$

\vspace{5pt}
\noindent
Here\hfill $T_{\cal D}$\hfill is\hfill the theory\hfill defined\hfill in Section\hfill \ref{subsec:FormalTheoryD},\hfill the meaning\hfill of\hfill the predicate\hfill $K(\texttt{\tt e{\rdot}xs}, \texttt{\tt e{\rdot}ys})$\hfill is\hfill that\\  
\texttt{e{\rdot}xs} and \texttt{e{\rdot}ys} are two consecutive Fibonacci words. Negation of the last two axioms 
corresponds to the properties defined by the predicates \texttt{A}, \texttt{B} given in Example \ref{ex:FibTestOpen}.
Stepwise computation of a given Fibonacci word \texttt{e{\rdot}xs$_0$} corresponds to stepwise derivability of  $\exists \texttt{\tt e{\rdot}ys}\, (K(\texttt{\tt e{\rdot}xs}_0, \texttt{\tt e{\rdot}ys}))$.
Mace4 is able to refute 
$\exists \texttt{\tt e{\rdot}xs}\, \exists \texttt{\tt e{\rdot}ys}\, (K(\texttt{\tt e{\rdot}xs}, \texttt{\tt e{\rdot}ys}) \wedge B(\texttt{\tt e{\rdot}xs}))$ and  
$\exists \texttt{\tt e{\rdot}xs}\, \exists \texttt{\tt e{\rdot}ys}\, (K(\texttt{\tt e{\rdot}xs}, \texttt{\tt e{\rdot}ys}) \wedge A(\texttt{\tt e{\rdot}xs}))$, 
by finding countermodels $M_{1}$ and $M_{2}$  of sizes 5 and 11, respectively.
Description of these models can be found in \cite{Lisitsa:TalkCachan12}.

\section{Regular Invariants and Beyond}\label{sec:IrregularExamples}

The finite models produced above can be seen as  compact representations of the \emph{regular} invariants  (see Section \ref{subsec:Invariants}) and \cite{Lisitsa:RTA12}) sufficient to prove 
 safety, i.e.,  unreachability properties.  
 The example above shows that enhancing of the supercompilation 
 by the ``regular verifying power'' of FCM may be beneficial for producing non-trivial program transformations. 
What is interesting here is that the mechanisms for program  analysis and transformation deployed within  supercompilation do  not cover all the regular power of FCM but  may go  beyond that. 

Given a program rule $l = r$, obviously, using an \texttt{e/t}-variable $v$ such that $\mu_v(r) > 1$ may lead to one-step computing a non-regular formal language ${\cal H} \subset {\cal D}$. Such a language ${\cal H}$ may also be generated by recursion. The following two examples deal with 
 that case.
The examples (being variations of the rules borrowed from \cite{BoichutHeam:TheoreticalLimit}) define the empty partial function. The programs $\langle \tau, R \rangle$ below 
never reach their exits from recursions. 
The exits are defined in the first two  rules (in both programs). 
The first recursion given in the program 
$F$ 
constructs two equal strings in the second and third arguments, using 
 associativity of concatenation.
 Respectively  the second recursion given in the program $G$ constructs two equal binary trees, using the parenthesis constructor. 
The first arguments of the programs are the recursion depths. 
Evaluation of the programs generates  respectively the following formal languages of terms:\\  
$H_f = $ \texttt{\tt f($K$,'h$^n$'\,{\rm :}'A', 'h$^n$'\,{\rm :}'A')}, 
$H_g = $ \texttt{\tt g($K$,$\overbrace{\texttt{\tt ('h'\,{\rm :}}}^{n}$'A'$\overbrace{\texttt{\tt )}}^{n}$, $\overbrace{\texttt{\tt ('h'\,{\rm :}}}^{n}$'A'$\overbrace{\texttt{\tt )}}^{n}$)},\\
where $K = ($\texttt{\tt 'b' | 'c'}$)^m$ and $m \in \Nat$.

\begin{example}\label{ex:f} 
The program $F$ is $\langle \tau, R \rangle$, where
$\tau$ is \texttt{\tt f(e{\rdot}ps, 'A', 'A')} and $R$ is\\
\texttt{\tt f($\empt$, 'h'\,{\rm :}\,e{\rdot}xs, 'A') = 'A';}\\
\texttt{\tt f($\empt$, 'A', 'h'\,{\rm :}\,e{\rdot}ys) = 'A';}\\
\texttt{\tt f('b'\,{\rm :}\,e{\rdot}ps, e{\rdot}xs, e{\rdot}ys) = f(e{\rdot}ps, 'h'\,{\rm :}\,e{\rdot}xs, 'h'\,{\rm :}\,e{\rdot}ys);}\\
\texttt{\tt f('c'\,{\rm :}\,e{\rdot}ps, 'h'\,{\rm :}\,e{\rdot}xs, 'h'\,{\rm :}\,e{\rdot}ys) = f(e{\rdot}ps, e{\rdot}xs, e{\rdot}ys);}
\end{example}

\begin{example}\label{ex:g}
The program $G$ is $\langle \tau, R \rangle$, where
$\tau$ is \texttt{\tt g(e{\rdot}ps, 'A', 'A')} and $R$ is\\
\texttt{\tt g($\empt$, ('h'\,{\rm :}\,e{\rdot}xs), 'A') = 'A';}\\
\texttt{\tt g($\empt$, 'A', ('h'\,{\rm :}\,e{\rdot}ys)) = 'A';}\\
\texttt{\tt g('b'\,{\rm :}\,e{\rdot}ps, e{\rdot}xs, e{\rdot}ys) = g(e{\rdot}ps, ('h'\,{\rm :}\,e{\rdot}xs), ('h'\,{\rm :}\,e{\rdot}ys));}\\
\texttt{\tt g('c'\,{\rm :}\,e{\rdot}ps, ('h'\,{\rm :}\,e{\rdot}xs), ('h'\,{\rm :}\,e{\rdot}ys)) = g(e{\rdot}ps, e{\rdot}xs, e{\rdot}ys);}
\end{example}

Proving the emptiness of the partial functions defined by $F$ and $G$ can be seen as safety verification, that is  unreachability of  the first two rules of  $F$ and $G$ (i.e., the exits from the recursions). In \cite{BoichutHeam:TheoreticalLimit} Y. Boichut and P.-C. Heam showed that this safety property cannot be proved by safety verification techniques using \emph{ regular invariants}.
It means in particular that FCM won't help in proving that. 
On the other hand well-known specialization methods can prune away the recursion exits from 
 program $G$.  
Indeed, the configuration sequence on the recursion path produced by the unfolding algorithm and outgoing from the initial configuration {\small{ \texttt{g(e{\rdot}ps,'A','A')} }} is: {\small{ \texttt{g(e{\rdot}ps,'A','A')}, \texttt{g(e{\rdot}ps$_1$,('h\,'\conct\,'A'),('h\,'\conct\,'A'))},  \texttt{g(e{\rdot}ps$_2$,('h\,'\conct ('h\,'\conct\,'A')),('h\,'\conct ('h\,'\conct\,'A')))},}} $\ldots$ Generalization algorithms based on the Higman-Kruskal relation \cite{Kruskal} will construct one of the following configurations: {\small{ \texttt{g(e{\rdot}ps,t{\rdot}x,t{\rdot}x)},\\
 \texttt{g(e{\rdot}ps$_1$,('h\,'\conct t{\rdot}x),('h\,'\conct t{\rdot}x))}, \texttt{g(e{\rdot}ps$_2$,('h\,'\conct ('h\,'\conct t{\rdot}x)),('h\,'\conct ('h\,'\conct t{\rdot}x)))},}} $\ldots$\\ 
Note that \texttt{t{\rdot}x} here denotes a standard variable like those occurring in Lisp-like languages to denote lists. 
Now obviously, the exit branches from the recursion will be pruned away from the unfolding tree. 

The associative case used in $F$ is more difficult. The supercompiler SCP4 recognizes the emptiness of 
 both of the partial functions: 
supercompiling 
 program $G$ results in a message on the empty partial function, while supercompiling 
 program $F$ results  in a program without  \emph{syntactic} exits from recursions.

Thus a finite countermodel finder and a specializer have incomparable 
power for verification and analysis and their joint use may be useful.

\section{Conclusions and Future Work}

 External provers were used in automated program   transformers    including specializers for a long time. For instance, in 1988 Y.~Futamura used such an external tool for proving some properties of parameterized   configurations  
  of programs  being specialized   \cite{Futamura:2002, Futamura:1988}. 
Moreover, 
  given a program specializer written in a language ${\cal U}$, ${\cal U}$ may include non-trivial semantics mechanisms in itself. 
  The mechanisms may allow us to implement non-trivial basic program analysis directly by means of the ${\cal U}$-semantics, 
  i.e., through a \emph{local} syntactical construction 
  without intricate programming. 
  Such 
  mechanisms may be considered as external tools with respect to the specializer. Examples of such programming languages are Prolog and Refal \cite{Turchin:Refal5}. For example, F. Fioravanti, A. Pettorossi and M. Proietti \cite{FioPettPro:01, FioPettProSen:04, FioPettProSen:11}, as well as several other authors, develop an unfold-fold based transformation technique for constraint logic programs with negation, implementing their transformers using constraint logic programming. 
 
The use of countermodels for the execution and analysis of logic programs has been considered in
the paper \cite{BVWD:2000}. It has been noticed that the failure of a query $Q$ for a logic program $P$  can be established by finding a countermodel for $P \rightarrow Q$. Furthermore a particular strategy using pre-interpretations (i.e., interpretations of the predicate symbols only, ignoring constructors and data) combined with the use of an abduction mechanism is proposed and compared with unrestricted search of countermodels.
Such a technique can be adapted for term rewriting systems and functional languages.


The approach we presented in this paper is related also to the work on \emph{abstract interpretations} 
\cite{Cousot:1977,Leuschel:99} and \emph{regular types} \cite{JBS:2011}.
The work  \cite{GH:2004} explicitly connects both areas
and demonstrates the transformations of the  set of regular type definitions corresponding to the finite tree automata,   into a finite pre-interpretation for a logic program, which is then 
 used for 
 program analysis and verification. The core of the transformation is a determinization procedure for a non-deterministic tree automata. 
 The difference with our approach, apart of obvious differences between logic and functional programming languages considered,  is that  \cite{GH:2004} deals with specific approach for pre-interpretation building, while we abstract away the details 
of  the model building procedure, 
which is used as an oracle. 
Still after appropriate translations the approach of \cite{GH:2004} can be used  for the tasks considered in this paper and we plan to explore this issue  in  the future work. 

In this paper we have shown that integrating 
a finite countermodel finder in a supercompiler may provide new features for non-trivial program transformations, which in turn may be used for non-trivial verification of safety properties of
 programs.
In particular, global unreachability of some new kind of regular formal languages 
constructed over the system states sometimes may be recognized. 
 Furthermore, Examples \ref{ex:f} and \ref{ex:g} demonstrate that 
Finite Countermodel Method (FCM) may be strengthened by 
supercompilation tools. As regards to this matter we would like to refer to an interesting example given by the researchers mentioned above,  working in the context of Prolog \cite{FioPettProSen:04}. They derive a one-counter machine from a constrained regular language specification. The corresponding residual program tests that  a string of a given length $n$ 
does not belong to the language $\{ \texttt{'a}^m\texttt{'\,\conct\,'b}^n\texttt{'} |\, m = n \ge 0 \}$.

Above we have described just the first steps and experiments in integrating FCM in a supercompiler. The examples given in Section \ref{sec:IrregularExamples} motivate future development of a compiler from a functional language (in our case, Refal)  to a first-order logic language.
The compiler should protect as many syntax properties of the program $\langle \tau, R \rangle$ being compiled as possible.
The \emph{overapproximated reachability} of the $\langle \tau, R \rangle$ states from $\tau$ should correspond to \emph{derivability} in the corresponding compiled program.
We conclude with the following note: FCM may be used for recognizing unreachable intermediate subgraphs generated by supercompilation even if the subgraphs are not self-sufficient (see Section \ref{sec:GlobalUnreachability}). 
In such cases we have to consider the external functions to be unknown.


\section*{Acknowlegements} 

We are grateful to the reviewers of the paper for their generous and constructive comments, which 
allowed us to improve the presentation of this paper and gives us lines for future work.

\label{sect:bib}
\bibliographystyle{plain}

\bibliography{lisitsa_nemytykh}

\end{document}